# IMPURITY-IMPURITY PAIR CORRELATION FUNCTION AND PARAMAGNETIC-TO-FERROMAGNETIC TRANSITION IN THE RANDOM ISING MODEL

*D. Ivaneyko[1], B. Berche[2], Yu. Holovatch[3], Ja. Ilnytskyi[4]*

[1]*Ivan Franko National University of Lviv, 79005, Lviv, Ukraine*
e-mail: ivaneiko@ktf.franko.lviv.ua

[2]*Laboratoire de Physique des Matěriaux, Universitě Henri Poincaru, Nancy 1, 54506 Vandњuvre les Nancy Cedex, France*
e-mail: berche@lpm.u-nancy.fr

[3]*Institute for Condensed Matter Physics, National Acad. Sci. of Ukraine, 79011 Lviv, Ukraine*
*Institut fьr Theoretische Physik, Johannes Kepler Universitдt Linz, 4040 Linz, Austria*
e-mail: hol@icmp.lviv.ua

[4]*Institute for Condensed Matter Physics, National Acad. Sci. of Ukraine, 79011 Lviv, Ukraine*
e-mail: iln@icmp.lviv.ua

In this Monte Carlo study we concentrated on the influence of non-magnetic impurities arranged as the lines with random orientation on paramagnetic-to-ferromagnetic phase transition in the *3D* Ising model. Special emphasize is given to the long-distance decay of the impurity-impurity pair correlation function. It is shown that for the lattice sizes considered ($L$=10-96) and for the two different impurity distributions (purely random and mutually avoiding lines) the function is governed by the power law of $1/r^a$ with an universal exponent $a≈2$. This result supports our findings about the numerical values of the critical exponents governing magnetic phase transition in the *3D* Ising model with long-range-correlated disorder.

The ferromagnetic phase transition in the three-dimensional (*3D*) Ising model with quenched long-range-correlated impurities is governed by the critical exponents that are different of both those for the pure *3D* Ising model and for the *3D* Ising model with uncorrelated impurities [1-5]. In particular, we are interested in the case, when the impurity-impurity pair correlation function $h(r)$ decays at large separations $r$ accordings to a power law [1]:

$$h(r) \approx 1/r^a, \qquad r \to \infty. \qquad (1)$$

Besides purely academic interest, a reason behind such a choice is that the power law decay (1) allows a direct geometrical interpretation. Indeed, for integer $a$ it corresponds to the lines (at $a=D-1$) or the planes (at $a=D-2$) of impurities of random orientation [1]. Moreover, non-integer $a$ is sometimes treated in terms of impurities fractal dimension [6].

Whereas both analytical and numerical studies agree on the fact that the *3D* Ising model with long-range correlated disorder (i.e. when $a<D$) possesses a new universality class [1-5], the numerical values of the critical exponents are found to be rather different. Indeed, the renormalization group (RG) estimates of Ref. [1] gave the values of the exponents in the first order of $\varepsilon=4-D$, $\delta=4-a$-expansion. Furthermore, the first-order result for the correlation length critical exponent $\nu=2/a$ was conjectured to be an exact one [1]. On contrary, the second-order RG calculations of Ref. [2] lead to non-trivial dependence of exponents on the correlation parameter $a$. MC simulations also split into two groups. First, the random Ising model with power-law correlated non-magnetic impurities was considered in Ref. [3]. There, the impurities were simulated in two different ways: (i) as point-like particles, correlated according to (1) with correlation exponent $a=2$ and (ii) as the lines of random orientation. An outcome of the simulations favoured theoretical predictions of Ref. [1]. Indeed, for the impurity concentration $p$=0.2 the correlation length and pair correlation function exponents were estimated by means of combination of Wolff and Swendsen-Wang algorithms as $\nu$=1.012(10) and $\eta$=0.043(4) [3], whereas the theoretical estimate of Ref. [1] reads: $\nu(a=2)=1$, $\eta=0$. However, the following MC simulation [4] questioned results of Refs. [1,3]. Two sets of estimates for the exponents obtained there by using different algorithms at $p$=0.2 read: $\nu$=0.719(22), $\beta$=0.375(45) (short-time critical dynamics with Metropolis algorithm), and $\nu$=0.710(10), $\gamma$=1.441(15), $\beta$=0.362(20) (finite-size scaling with Wolff algorithm). In turn, these results support a theoretical estimate of Ref. [2]: $\nu$=0.7151 (note however, that $\eta$= -0.0205 in [2]). As a possible reason for the discrepancy with Refs. [1,3] the authors of Ref. [4] mention that they implemented a mutual avoidance condition on the lines of impurities, whereas it was not the case in the simulations of Ref. [3].

To resolve such a bias, we performed MC simulations of the *3D* Ising model with the impurities arranged as lines of random orientation [5]. Our estimates for the exponents differ from the results of the two numerical simulations performed so far [3,4] and are in favour of a non-trivial dependency of the critical exponents on the peculiarities of long-range correlations. Moreover, we have analysed both previously considered cases of purely random and mutually avoiding impurity lines distributions. No difference was found within the error bars (see below for more details).

One more question remained unanswered in the above-mentioned context. Namely, in the numerical simulations performed so far [3-5] it was tacitly assumed that the impurity-impurity pair correlation function power law asymtotics holds for the randomly oriented impurity lines with an exponent $a=2$. Although, it is certainly true for an infinite system, it is not obvious that such behaviour holds for the finite-size systems considered during simulations. Therefore, the goal of this paper is to supply the MC simulations of the phase transition in magnetic subsystem by simultaneous control of the structural properties of the impurities.

We consider a *3D* Ising model with non-magnetic sites arranged in a form of randomly oriented lines. The Hamiltonian reads:

$$H = -J \sum_{\langle ij \rangle} c_i c_j S_i S_j, \qquad (2)$$

where the summation is over the nearest neighbour sites of a s.c. lattice of linear size $L$, $J>0$ is the interaction constant, Ising spins $S_i = \pm 1$, and $c_i = 0,1$ is the occupation number for the *i*-th site. Non-magnetic sites ($c_i = 0$) are located along the lines and quenched in a fixed configuration. To ensure an isotropic distribution of lines, we take their number to be the same along each axis.

We performed the MC simulations by means of the Wolff cluster algorithm [7] using histogram reweighting technique [8] imposing periodic boundary conditions, measuring system magnetisation, energy, Binder cumulant, and magnetic susceptibility at the critical temperature for the lattices of varying sizes $L=10-96$ and applying finite-size scaling technique to extract the values of the critical exponents. The impurity concentration was taken to be $p=0.2$ both to adhere previous MC simulations [3,4] as well as to minimize possible correction-to-scaling effects. The presence of quenched disorder leads to two different types of averaging to be performed: besides the Botzmann averaging, the observables are to be averaged with respect to different disorder realizations. To perform the averaging over different disorder configurations we generated $10^4$ lattice samples for the sizes $L=10-48$ and $10^3$ samples for $L=64,96$. To accomplish the Boltzmann averaging for each disorder realization, a run of $250\tau_E$ Monte Carlo steps (MCS) was performed for system equilibration with following 20000 MCS for further calculations at $L=10-32$. For the larger lattice sizes, $L=48-96$ the number of MCS steps was $10^3\tau_E$ ($\tau_E$ being energy autocorrelation time) [9]. Further details of our simulations are reported elsewhere [5].

As has already been outlined in the introduction, we are interested in the analysis of two different variants of non-magnetic sites distribution: in the first one the lines of impurities are randomly oriented and some of them may intersect (from now on we call such situation 'distribution ***A***'); whereas in the second variant we impose a mutual avoidance condition on the randomly oriented lines ('distribution ***B***', correspondingly). Theoretically, these two distributions may lead to different critical behaviours, as far as only the distribution ***A*** corresponds to the impurities in the form of lines, whereas the distribution ***B*** may result in objects of a different dimension. However it is not obvious that such effect may show up and cause any influence on the critical behaviour for the systems of sizes considered in the simulation. Indeed, the results for the critical exponents we obtain in the simulations for the above two distributions read [5]:

***A***: $\nu=0.864(10)$, $\beta=0.519(11)$, $\gamma=1.555(26)$; (3)

***B***: $\nu=0.872(19)$, $\beta=0.522(16)$, $\gamma=1.450(39)$. (4)

The exponents $\nu$ and $\beta$ obtained do agree within the confidence interval whereas exponent $\gamma$ differs within 3%. Already the above numbers enabled us to arrive to the conclusion [5] that the difference between the results of simulations performed in Ref. [3] and Ref. [4] can not be caused solely by the difference between impurity line distributions ***A*** and ***B***.

To complete an analysis of the paramagnetic-to-ferromagnetic phase transition that has led to the estimates (3), (4) for the critical exponents, we studied the behaviour of the impurity-impurity pair corelation function $h(r)$ for the distributions ***A*** and ***B*** for the lattices considered. To this end, we define $h(r)$ in terms of the radial distribution function $g(r)$

$$h(r) = g(r) - 1, \qquad (5)$$

where $g(r)$ is given by [10]:

$$g(r) = \frac{\langle n(r) \rangle}{\frac{4}{3}\pi((r+\delta r)^3 - r^3)}. \qquad (6)$$

In (6), $\langle n(r) \rangle$ means average number of non-magnetic sites which lie within distance $\delta r$ of a sphere of radius $r$ that contains a non-magnetic site in the origin. Note, that for the diluted system (6) is to be normalized by the non-magnetic component concentration $p$.

By applying Eq. (6) for each disorder realization it is straightforward to find the impurity-impurity pair correlation function by counting the number of non-magnetic sites that lie within the distance interval $[r, r+\delta r]$ from the given one. Then, the resulting histogram is to be averaged over different disorder realizations (different samples). The number of samples was taken $N=1000$ for the lattice sizes $L=10-64$ and $N=100,50$ for the lattices with *L*=96,128, respectively. Note, that for a given sample the statistics is enriched by placing in turn each non-magnetic site at the origin. This increases the number of data points for the largest lattice sizes *L*=128 by $pL^3 \approx 10^5$. In Figs. 1, 2 we give a typical outcome of the analysis: configurationally averaged impurity-impurity pair correlation function *h(r)* for one of the lattice sizes (*L*=96) and for the impurity distributions *A* and *B*. Fit to a power-law (1) is shown by a solid line.

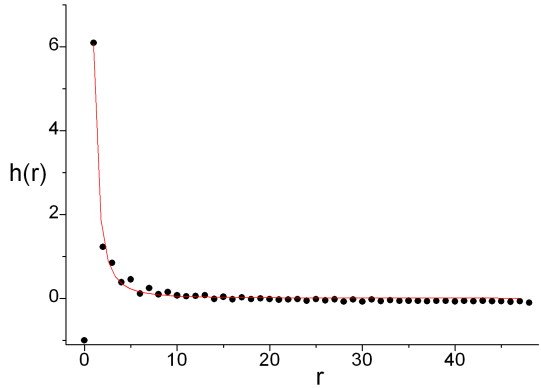

***Fig.1.*** *Impurity-impurity pair correlation function h(r) for the distribution **A** and lattice size L=96. Solid line (red online) shows a power-law fit (1) with an exponent a=2.05 (see Table 1)*

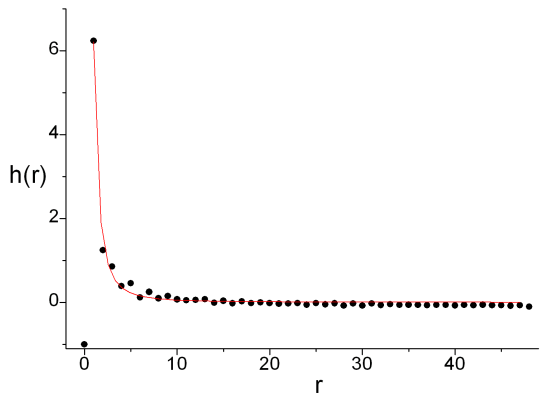

***Fig.2.*** *Impurity-impurity pair correlation function h(r) for the distribution **B** and lattice size L=96. Solid line (red online) shows a power-law fit (1) with an exponent a=2.06 (see Table 1)*

It is appropriate to make several comments before passing to numerical estimates of the exponents governing such fits. For the sake of theoretical analysis, one is interested in the behaviour of *h(r)* for large *r*. In the simulations, *r* is limited by the linear lattice size *L*. However, because of the system finite size and a presence of periodic boundary conditions, the cutoff distance for the interparticle correlations is *L/2*. Therefore, although the theoretical estimates of Refs. [1,2] are intended for the systems where (1) holds in asymptotics, in the finite-size systems with randomly distributed lines the power-law decay (1) is observed in the wide region of distances, starting already from *r* = 1 (note that at the origin *h(r)* = -1 by definition (5)).

Table 1 summarizes our results of the fits of the impurity-impurity pair correlation function *h(r)* to the power law for lattice sizes L = 10-128 and for impurity lines distributions *A* and *B*. There are several conclusions that follow from these data:

(i) the power-law behaviour of *h(r)* is confirmed for the lattice sizes used in the simulations;

(ii) distributions *A* and *B* are characterised by the same (within the confidence interval) value of the exponent *a*;

(iii) the value of the exponent *a* within the confidence interval does not vary with *L*.

The last fact is of crucial importance as far as it justifies application of the finite-size-scaling technique used to extract the values of critical exponents that govern paramagnetic-to-ferromagnetic phase transition. Indeed, if *a* were found to vary with *L* an application of this technique would be questioned, as far as in this case one faces an analysis of systems which are differently correlated for each *L*. Note also that the correlation parameter is $a \approx 2$, and the deviations found are of order of several percents and do not suffice to cause changes in the critical exponents values within the confidence interval of data given in (3), (4).

***Table 1.*** *The values of exponent a for different lattice sizes L and for two different impurity lines distributions. Distribution A: lines are allowed to intersect; distribution **B**:  mutually avoiding lines*

| L | Distribution *A* | Distribution *B* |
|---|---|---|
| 10 | 2.16 ± 0.18 | 2.17 ± 0.20 |
| 12 | 2.08 ± 0.17 | 2.10 ± 0.18 |
| 16 | 2.07 ± 0.13 | 2.08 ± 0.14 |
| 24 | 2.04 ± 0.10 | 2.06 ± 0.10 |
| 32 | 2.03 ± 0.08 | 2.06 ± 0.10 |
| 48 | 2.04 ± 0.07 | 2.05 ± 0.08 |
| 64 | 2.04 ± 0.06 | 2.05 ± 0.06 |
| 96 | 2.05 ± 0.06 | 2.06 ± 0.06 |
| 128 | 2.05 ± 0.06 | 2.06 ± 0.06 |

Together with the results for critical exponents (3), (4) the data presented in Table 1 bring about the fact that the *3D* Ising model with long-range-correlated disorder in the form of  non-magnetic impurity lines of random orientation belongs to the new universality class. The values of the exponent obtained certainly differ from those of the pure *3D* Ising model ($\nu$ = 0.630(1), $\beta$ = 0.3265(15), $\gamma$ = 1.237(3) [11]) as well as from the *3D* Ising model with uncorrelated impurities ($\nu$ = 0.68(2), $\beta$ = 0.35(1), $\gamma$ = 1.34(1) [12]). However, our results differ from previous MC estimates [3,4] of the exponents for *3D* Ising model with long-range-correlated disorder (see the numbers given at the beginning of this paper). The reason for the discrepancy remains unclear.  Furthermore, for the lattice sizes considered the constraint of mutual avoidance imposed on the impurity lines appears to be an irrelevant one.

Results discussed here were presented at the 2nd International Conference on Quantum Electrodynamics and Statistical Physics (Kharkiv, 19-23 September, 2006). Yu. H. deeply acknowledges Yurij Slyusarenko for his kind hospitality during stay in Kharkiv.## REFERENCES

1. A. Weinrib, B.I. Halperin. Critical phenomena in systems with long-range-correlated quenched disorder // *Phys. Rev. B*. 1983, v. 27, p. 413-427.

2. V.V. Prudnikov, P.V. Prudnikov, A.A. Fedorenko. Field-theory approach to critical behavior of systems with long-range correlated defects // *Phys. Rev. B*. 2000, v. 62, p. 8777-8786.

3. H.G. Ballesteros, G. Parisi. Site-diluted three-dimensional Ising model with long-range correlated disorder // *Phys. Rev. B*. 1999, v. 60, p. 12912-12917.

4. V.V. Prudnikov, P.V. Prudnikov, S.V. Dorofeev, V.Yu. Kolesnikov. Monte Carlo studies of critical behaviour of systems with long-range correlated disorder // *Condens. Matter Phys.* 2005, v. 8, p. 213-224.

5. D. Ivaneyko, B. Berche, Yu. Holovatch, J. Ilnytskyi. *On the universality class of the 3d Ising model with long-range-correlated disorder.* cond-mat/ 0611568, 2006, 25 p. (submitted to Physica A).

6. C. Vasquez, R. Paredes, V.A. Hasmy, R. Jullien. New universality class for the three-dimensional XY model with correlated impurities: application to $^4$He in aerogels // *Phys. Rev. Lett.* 2003, v. 90, p. 70602.

7. U. Wolff. Collective Monte Carlo updating for spin systems // *Phys. Rev. Lett.* 1989, v. 62, p. 361-364.

8. A.M. Ferrenberg, R.H. Swendsen. New Monte Carlo technique for studying phase transitions // *Phys. Rev. Lett.* 1988, v. 61, p. 2635-2638.

9. D. Ivaneyko, J. Ilnytskyi, B. Berche, Yu. Holovatch. Local and cluster critical dynamics of the 3d random-site Ising model // *Physica A*. 2006, v. 370, p. 163-78.

10. P.M. Chaikin, T.C. Lubensky. *Principles of Condensed Matter Physics.* Cambridge University Press, 1997.

11. H.W.J. Blцte, E. Luijten, J.R. Heringa. Ising universality in three dimensions: a Monte Carlo study // *J. Phys. A.* 1995, v. 28, p. 6289-6313.

12. P.-E. Berche, C. Chatelain, B. Berche, W. Janke. Bond dilution in the 3D Ising model: a Monte Carlo study // *Europ. Phys. J. B.* 2004, v. 38, p. 463-474.
**ПАРНАЯ КОРЕЛЯЦИОННАЯ ФУНКЦИЯ ПРИМЕСЬ-ПРИМЕСЬ И ПЕРЕХОД ПАРАМАГНЕТИК-ФЕРРОМАГНЕТИК В НЕУПОРЯДОЧНЕНОЙ МОДЕЛИ ИЗИНГА**

*Д. Иванейко, Б. Берш, Ю. Головач, Я. Ильницкий*В роботе обсуждаются результаты Монте Карло иследований влияния немагнитных примесей в виде линий со случайной ориентацией на фазовый переход парамагнетик-ферромагнетик в трехмерной модели Изинга. Особое внимание уделено угасанию на больших расстояниях парной корреляционной функции примесь-примесь. Для обсуждаемых размеров решеток ($L$=10-96) и для двух типов распределений примесей (пересекающиеся и непересекающиеся линии) показано, что функция подчиняется закону $1/r^a$ с универсальным показателем $a \approx 2$. Этот результат поддерживает полученные нами ранее данные о числовых значениях критических показателей магнитного фазового перехода в трехмерной модели Изинга со скоррелированным на больших расстояниях беспорядком.**ПАРНА КОРЕЛЯЦІЙНА ФУНКЦІЯ ДОМІШКА-ДОМІШКА І ПЕРЕХІД ПАРАМАГНЕТИК-ФЕРОМАГНЕТИК В НЕВПОРЯДКОВАНІЙ МОДЕЛІ ІЗИНГА**

*Д. Іванейко, Б. Берш, Ю. Головач, Я. Ільницький*В роботі обговорюються результати досліджень методом Монте Карло впливу немагнітних домішок у вигляді ліній з випадковою орієнтацією на фазовий перехід парамагнетик-феромагнетик в тривимірній моделі Ізинга. Особливу увагу приділено загасанню на великих відстанях парної кореляційної функції домішка-домішка. Для розглянених размірів граток ($L$=10-96) і для двох типів розподілу домішків (лінії, що перетинаються і лінії, що не перетинаються) показано, що функція має степеневий вигляд $1/r^a$ з універсальним показником $a \approx 2$. Цей результат підтверджує отримані нами числові значення критичних показників магнітного фазового переходу в тривимірній моделі Ізинга з далекосяжно-скорельованим безладом.4